\documentclass[aps,preprint]{revtex4}%
\usepackage{amsfonts}%
\usepackage{amsmath}%
\usepackage{amssymb}%
\usepackage{graphicx}

\begin{document}
\title{Effect of the cosmological constant in the Hawking radiation of 3D charged
dilaton black hole }
\author{I.Sakalli}
\affiliation{AS 345, Physics Department, Eastern Meditterranean University, Gazimagosa,
North Cyprus via Mersin 10, Turkey}
\pacs{PACS number}

\begin{abstract}
This paper deals with the semiclassical radiation spectrum of static and
circularly symmetric 3D charged dilaton black holes with cosmological constant
$\Lambda$ in non-asymptotically flat spacetimes. We first review the 3D
charged dilaton black holes which are solution to low-energy string action.
The wave equation of a massless scalar field is shown to be exactly solvable
in terms of hypergeometric functions. Thus, the radiation spectrum and its
corresponding temperature are obtained, precisely. Computations at high
frequency regime show that the radiation spectrum yields the Hawking
temperature of the black hole with no charge. Unlike the chargeless case, the
Hawking temperature of the charged dilaton black holes is derived from the
radiation spectrum at the low frequencies. The utmost importance of the
$\Lambda$ in the latter result is highlighted.

\end{abstract}
\maketitle

\section{Introduction}

Since the seminal works of \cite{hawking74,hawking75,hawking76} and
\cite{bekenstein73}, it is well understood that a black hole (BH) can emit
particles from its event horizon with a certain temperature, so-called the
Hawking temperature ($T_{H}$), which is proportional to the surface gravity,
$\kappa,$ of the BH. Therefore, a distant observer could detect a thermal
radiation, i.e. Hawking radiation, that spreads throughout the BH
surroundings. In fact, this is the way that a BH ends. Today there are various
methods to compute the $T_{H}$ in the literature, (see for instance
\cite{damour76,gibbons77,strominger96,srinivasan99,angheben05,robinson05,banerjee08}%
, and it is always high on the agenda to find alternative derivations. As for
us, none of the methods is catchy as the Hawking's original study
\cite{hawking75,hawking76} in which the Hawking temperature is revealed by the
computation of the Bogoliubov coefficients between in and out states for a
collapsing BH. Another alternative method which is conclusive as the method of
the Bogoliubov coefficients is the computation of the reflection coefficient
of an incident wave by the BH. Thereby, the thermal radiation spectrum and
hence the $T_{H}$ are obtained precisely \cite{clement07}. The latter method
especially serves the purpose when the wave equation on the associated
geometry is solved exactly. This method is known as the "semi-classical
radiation spectrum method", and we call it as \textit{SCRSM }throughout the
paper like in \cite{mazhari09}.

Recent studies \cite{clement07,mazhari09} have shown us that when the
\textit{SCRSM} is applied for the non-asymptotically flat (NAF) charged
massive dilaton BHs without cosmological constant $\Lambda$ in which they have
only a single (event) horizon, one can reach to the $T_{H}$ which is in
general computed by the conventional method: $T_{H}=\frac{\kappa}{2\pi}$. The
way followed so far for this purpose in the \textit{SCRSM }was to take merely
the high frequency limit of the obtained temperature $T(\omega)$ of the
radiation spectrum in which $\omega$\ represents the frequency of the scalar
wave. The main aim of this paper is to explore the effect of the $\Lambda$\ on
the radiation spectrum of the charged dilaton BHs. Namely, we intend to fill a
gap in this line of thought. To this end, we will consider another NAF static
and charged dilaton BH. Unlike the previous studies, the considered BH is a
three dimensional (3D) solution to the Einstein-Maxwell-Dilaton theory with
cosmological constant $\Lambda$\ (EMD$\Lambda$ theory), which is found by
\cite{chan94}. As implied above, the inclusion of the $\Lambda$ enables the BH
to possess two horizons instead of one. Here, after a straightforward
calculation of the \textit{SCRSM}, it is seen that the high frequency limit of
the computed $T(\omega)$ does not lead it directly to the $T_{H}$ of the
charged dilaton BHs as such in the former studies \cite{clement07,mazhari09}.
For matching, it is not only enough to consider the low frequency regime but
one needs\ to impose some special conditions on the wave as well. Another
interesting point which comes in this regard is that the $\Lambda$ remains as
a decisive factor on the associated conditions about the wave. All these
points together with supporting plots will be examined in the following sections.

The lay out of the paper is as follows: In Sect. II, we start with a brief
review of the 3D charged dilaton BH solution in the EMD$\Lambda$ theory. The
next section is devoted to the analytical computation of the $T(\omega)$ via
the \textit{SCRSM} for the associated BHs. Some remarks about how the
$T(\omega)$\ goes to the $T_{H}$ of the charged dilaton BH are made, and these
remarks are illustrated by the various plots.\ We draw our conclusions in
Sect. IV.

\section{Charged dilaton black holes in 3D}

In this section we will review the geometrical and thermodynamical properties
of the 3D charged dilaton BHs. The associated BHs were found long ago by
\cite{chan94}. These BHs are exact solutions to the equations of motion of the
EMD$\Lambda$ action which is conformably related to the low-energy string
action. This action is defined by%

\begin{align}
I &  =\int d^{3}x\sqrt{-g}\left(  \Re+2e^{b\phi}\Lambda-\frac{B}{2}\left(
\nabla\phi\right)  ^{2}\right.  \nonumber\\
&  \left.  -e^{-4a\phi}F_{\mu\nu}F^{\mu\nu}\right)  ,
\end{align}
where $F_{\mu\nu}$ stands for the Maxwell field, the constants $a,b,B$ are
arbitrary constants, $\phi$ is the dilaton field, $\Re$ is the Ricci scalar in
3D and $\Lambda,$\ as already mentioned in the previous section, represents
the cosmological constant; $\Lambda>0$ stands for the anti-de Sitter (AdS)
spacetime, and $\Lambda<0$ represents\ the de Sitter (dS) spacetime. The
corresponding static circularly symmetric solution \cite{chan94} to the above
action is given by%

\begin{equation}
ds^{2}=-f(r)dt^{2}+4\frac{r^{\frac{4}{N}-2}}{N^{2}\gamma\frac{4}{^{N}}%
}f(r)^{-1}dr^{2}+r^{2}d\theta^{2},
\end{equation}
where the metric function is%

\begin{equation}
f(r)=-Ar^{\frac{2}{N}-1}+\frac{8\Lambda r^{2}}{(3N-2)N}+\frac{8Q^{2}}{(2-N)N},
\end{equation}
One can readily see that the spacetime (2) possesses a NAF geometry, therefore
it does not behave asymptotically as dS or AdS \cite{chan94}. This is due to
the presence of the nontrivial dilaton field $\phi$. The corresponding dilaton
field $\phi$ and the non-zero maxwell tensor $F_{tr}$ are%

\begin{align}
\phi &  =\frac{2k}{N}\ln\left[  \frac{r}{\beta(\gamma)}\right]  ,\nonumber\\
F_{tr}  &  =\frac{2}{N}\gamma^{-\frac{2}{N}}r^{\frac{2}{N}-2}Qe^{4a\phi},
\end{align}
where $Q$ is related with electric charge, $\gamma$ and whence $\beta$ are
constants, and $k$ is another constant which is governed by%

\begin{align}
k &  =\pm\sqrt{\frac{N}{B}(1-\frac{N}{2})},\text{ \ \ \ \ }4ak=bk=N-2,\text{
}\nonumber\\
&  \therefore4a=b,
\end{align}
As it was stated in the paper \cite{chan94}, when $\left(  \Lambda,B\right)
>0$ and $2>N>\frac{2}{3}$ the solutions (2) and (3) represent BHs. In this
case, the constant $A$ of the metric function (3) is related with the
quasilocal mass $M$ \cite{brown93,liu03} of the BH. By employing the
Brown-York formalism \cite{brown93}, one can find the following relationship%

\begin{equation}
A=-\frac{2M}{N}.
\end{equation}
From the above expression, one can infer that for $M>0$, $A$ should take
negative values. Similar to the work \cite{ortega05}, throughout this paper we
restrict attention to the BHs described by $N=\gamma=\beta=a=\frac{b}{4}%
=\frac{B}{8}=-4k=1$. These choices modify the metric function, the dilaton
field and the Maxwell field to%

\[
f(r)=-2Mr+8\Lambda r^{2}+8Q^{2},
\]

\begin{equation}
\phi=-\frac{1}{2}\ln(r),
\end{equation}

\[
F_{tr}=\frac{2Q}{r^{2}}.
\]
When $M\geq8Q\sqrt{\Lambda}$, the spacetime (2) represents a BH. For
$M>8Q\sqrt{\Lambda}$, two horizons of the BH are found in the following locations%

\begin{equation}
r_{+}=M\frac{1+Y}{8\Lambda},\text{ \ \ \ }r_{-}=M\frac{1-Y}{8\Lambda},
\end{equation}
where%

\begin{equation}
Y=\sqrt{1-\frac{64Q^{2}\Lambda}{M^{2}}}\text{ \ \ \ \ \ \ \ }(0<Y<1),
\end{equation}
In (8), $r_{+}$\ and $r_{-}$ denote outer and inner horizons of the BH,
respectively. Besides, there is a timelike singularity at $r=0$. These charged
BHs possess a Hawking temperature ($T_{Hch}$) that it can be computed from the
definition given by \cite{wald84}%

\begin{align}
T_{H}  &  \equiv T_{Hch}=\frac{\kappa}{2\pi}=\frac{\left\vert g_{tt}^{\prime
}\right\vert }{4\pi}\left.  \sqrt{-g^{tt}g^{rr}}\right\vert _{r=r_{+}%
},\nonumber\\
&  =\frac{M}{4\pi r_{+}}Y,
\end{align}
The prime symbol in the foregoing equation, and in the following sections,
denotes the derivative with respect to the variable "$r$". For the extreme BHs
with $M=8Q\sqrt{\Lambda},$ which is equivalent to $Y=0,$ one can immediately
verify that $T_{Hch}$ vanishes. Furthermore, for the uncharged ($Q=0$ or
$Y=1$) BHs the Hawking temperature ($T_{Huch}$)\ becomes
\begin{equation}
T_{H}\equiv T_{Huch}=\frac{\Lambda}{\pi}.
\end{equation}

\section{Calculation of $T(\omega)$ via the \textit{SCRSM}}

In this section, by using the \textit{SCRSM }\cite{mazhari09}, we will make
more precise calculation of the temperature $T(\omega)$\ of the radiation for
the 3D charged dilaton BHs which are introduced in the previous section. After
obtaining the $T(\omega)$, its $T_{Huch}$ and $T_{Hch}$ limits will be
discussed in detail. For this purpose, we will first develop the wave equation
for a massless scalar field in the background of the charged dilaton BH.
Although the calculations made here share partial similarities with the
studies \cite{fernando05,fernando08}, for the sake of completeness we would
like to describe the details.

The general equation for a massless scalar field in a curved spacetime is
written as%

\begin{equation}
\square\Phi=0,
\end{equation}
where the d'Alembertian operator $\square$ is given by%

\begin{equation}
\square=\frac{1}{\sqrt{-g}}\partial_{\mu}(\sqrt{-g}g^{\mu\nu}\partial_{\nu}),
\end{equation}
The scalar wave function $\Phi$ of (12) can be separated into a radial
equation by letting%

\begin{equation}
\Phi=P(r)e^{-i\omega t}e^{\pm m\theta},
\end{equation}
where $m$ is a complex integration constant. Meanwhile, before proceeding to
the derivation of the radial equation, in order to facilitate the
computations, it is better to rewrite the metric function $f(r)$ (7) as \ %

\begin{equation}
f(r)=8\Lambda(r-r_{+})(r-r_{-}),
\end{equation}
After inserting the ansatz (14) into the wave equation (12), the radial
equation becomes%

\begin{equation}
f(r)P(r)^{\prime\prime}+f^{\prime}(r)P(r)^{\prime}+4\left(  \frac{r^{2}%
\omega^{2}}{f(r)}+m^{2}\right)  P(r)=0,
\end{equation}
The above equation can be solved in terms of hypergeometric functions. Here,
we give the final result as%

\begin{align}
P(r)  &  =C_{1}(r-r_{+})^{i\tilde{\omega}r_{+}}(r-r_{-})^{-i\tilde{\omega
}r_{-}}F\left[  \tilde{a},\tilde{b};\tilde{c};y\right] \nonumber\\
&  +C_{2}(r-r_{+})^{-i\tilde{\omega}r_{+}}(r-r_{-})^{-i\tilde{\omega}r_{-}}\\
&  \times F\left[  \tilde{a}-\tilde{c}+1,\tilde{b}-\tilde{c}+1;2-\tilde
{c};y\right]  .\nonumber
\end{align}
where $y=\frac{r_{+}-r}{r_{+}-r_{-}}$. The parameters of the hypergeometric
functions are%

\begin{align}
\tilde{a}  &  =\frac{1}{2}+i(\frac{\omega}{4\Lambda}+\sigma),\nonumber\\
\tilde{b}  &  =\frac{1}{2}+i(\frac{\omega}{4\Lambda}-\sigma),\\
\tilde{c}  &  =1+2i\tilde{\omega}r_{+},\nonumber
\end{align}
where%

\begin{align}
\sigma &  =\frac{1}{4\Lambda}\sqrt{\omega^{2}-\rho},\text{ \ \ }\rho
=4\Lambda\left(  \Lambda-2m^{2}\right)  ,\text{\ \ \ }\tilde{\omega}%
=\omega\lambda,\text{ }\nonumber\\
\lambda &  =\frac{1}{4\Lambda(r_{+}-r_{-})}.
\end{align}
and $\sigma$\ is assumed to have positive real values. Next, setting%

\begin{equation}
r-r_{+}=\exp(\frac{x}{\lambda r_{+}}),
\end{equation}
one gets the behavior of the partial wave near the outer horizon
($r\rightarrow r_{+}$) as%

\begin{equation}
\Phi\simeq C_{1}e^{i\omega(x-t)}+C_{2}e^{-i\omega(x-t)}.
\end{equation}
We assign the constants $C_{1}$ and $C_{2}$ as the amplitudes of the
near-horizon outgoing and ingoing waves, respectively.

To obtain the asymptotic solution ($r\rightarrow\infty$) of (17), one can
perform a transformation of the hypergeometric functions of any argument (let
us say $z$) to the hypergeometric functions of its inverse argument ($1/z$)
which is given as follows \cite{abram65}%

\begin{align}
F(\bar{a},\bar{b};\bar{c};z)  &  =\frac{\Gamma(\bar{c})\Gamma(\bar{b}-\bar
{a})}{\Gamma(\bar{b})\Gamma(\bar{c}-\bar{a})}(-z)^{-\bar{a}}\nonumber\\
&  \times F(\bar{a},\bar{a}+1-\bar{c};\bar{a}+1-\bar{b};1/z)\nonumber\\
&  +\frac{\Gamma(\bar{c})\Gamma(\bar{a}-\bar{b})}{\Gamma(\bar{a})\Gamma
(\bar{c}-\bar{b})}(-z)^{-\bar{b}}\\
&  \times F(\bar{b},\bar{b}+1-\bar{c};\bar{b}+1-\bar{a};1/z).\nonumber
\end{align}
Applying this transformation to (17), one can find the solution to the wave
equation in the asymptotic region as follows%

\begin{align}
\Phi &  \simeq(r-r_{-})^{-i\tilde{\omega}r_{-}}\left(  r-r_{+}\right)
^{-\frac{1}{2}}\nonumber\\
&  \times\left\{  D_{1}\exp i\left[  \frac{x}{\lambda r_{+}}(\sigma
+\tilde{\omega}r_{-})-\omega t\right]  \right. \\
&  \left.  +D_{2}\exp i\left[  \frac{x}{\lambda r_{+}}(-\sigma+\tilde{\omega
}r_{-})-\omega t\right]  \right\}  .\nonumber
\end{align}
On the other hand, since we consider the case of $r\rightarrow\infty,$ the
overall-factor term behaves as%

\begin{equation}
(r-r_{-})^{-i\tilde{\omega}r_{-}}\cong\exp i(-\frac{x\omega r_{-}}{r_{+}}),
\end{equation}
Therefore, the wave near the infinity (23) reduces to%

\begin{align}
\Psi &  \simeq\left(  r-r_{+}\right)  ^{-\frac{1}{2}}\left[  D_{1}\exp
i\left(  \frac{x}{\lambda r_{+}}\sigma-\omega t\right)  \right. \nonumber\\
&  \left.  +D_{2}\exp i\left(  -\frac{x}{\lambda r_{+}}\sigma-\omega t\right)
\right]  ,
\end{align}
where $D_{1}$ and $D_{2}$ correspond to the amplitudes of the asymptotic
outgoing and ingoing waves, respectively. One can derive the relations between
$D_{1}$, $D_{2}$ and $C_{1}$, $C_{2}$ as the following.%

\begin{equation}
D_{1}=C_{1}\frac{\Gamma(\tilde{c})\Gamma(\tilde{a}-\tilde{b})}{\Gamma
(\tilde{a})\Gamma(\tilde{c}-\tilde{b})}+C_{2}\frac{\Gamma(2-\tilde{c}%
)\Gamma(\tilde{a}-\tilde{b})}{\Gamma(\tilde{a}-\tilde{c}+1)\Gamma(1-\tilde
{b})},
\end{equation}

\begin{equation}
D_{2}=C_{1}\frac{\Gamma(\tilde{c})\Gamma(\tilde{b}-\tilde{a})}{\Gamma
(\tilde{b})\Gamma(\tilde{c}-\tilde{a})}+C_{2}\frac{\Gamma(2-\tilde{c}%
)\Gamma(\tilde{b}-\tilde{a})}{\Gamma(\tilde{b}-\tilde{c}+1)\Gamma(1-\tilde
{a})}.
\end{equation}
According to the \textit{SCRSM}, Hawking radiation is considered as the
inverse process of scattering by the BH in which the outgoing mode at the
spatial infinity should be absent \cite{clement07}. In short, $D_{1}=0$ in
(26). This yields the coefficient of reflection $R$ of the wave by the BH as%

\begin{equation}
R=\frac{\left\vert C_{1}\right\vert ^{2}}{\left\vert C_{2}\right\vert ^{2}%
}=\frac{\left\vert \Gamma(\tilde{c}-\tilde{b})\right\vert ^{2}\left\vert
\Gamma(\tilde{a})\right\vert ^{2}}{\left\vert \Gamma(1-\tilde{b})\right\vert
^{2}\left\vert \Gamma(\tilde{a}-\tilde{c}+1)\right\vert ^{2}},
\end{equation}
which corresponds to%

\begin{equation}
R=\frac{\cosh\left[  \pi\left(  \sigma-\frac{\omega}{4\Lambda Y}\right)
\right]  \cosh\left[  \pi\left(  \sigma-\frac{\omega}{4\Lambda}\right)
\right]  }{\cosh\left[  \pi\left(  \sigma+\frac{\omega}{4\Lambda Y}\right)
\right]  \cosh\left[  \pi(\sigma+\frac{\omega}{4\Lambda})\right]  },
\end{equation}
Since the resulting radiation spectrum is
\begin{align}
N  &  =\left(  e^{\frac{\omega}{T}}-1\right)  ^{-1},\nonumber\\
&  =\frac{R}{1-R}\rightarrow\text{ }T=T(\omega)=\frac{\omega}{\ln(\frac{1}%
{R})},
\end{align}
one can easily read the more precise value of the temperature as%

\begin{align}
T(\omega)  &  =\omega/\ln\left\{  \frac{\cosh\left[  \pi\left(  \sigma
+\frac{\omega}{4\Lambda Y}\right)  \right]  }{\cosh\left[  \pi\left(
\sigma-\frac{\omega}{4\Lambda Y}\right)  \right]  }\right. \nonumber\\
&  \left.  \times\frac{\cosh\left[  \pi(\sigma+\frac{\omega}{4\Lambda
})\right]  }{\cosh\left[  \pi\left(  \sigma-\frac{\omega}{4\Lambda}\right)
\right]  }\right\}  .
\end{align}
After analyzing the above result, it is seen that there are two prominent
cases depending on which frequency regime (high/low) is taken into consideration.

\textbf{Case I:} In the high frequency regime with the condition of
$\Lambda-2m^{2}\geq0$ (this case is also possible with a shift $m\rightarrow
im$ in the ansatz (14) of the wave function), the term $\frac{\omega}{4\Lambda
Y}\ $always predominates the$\ $physical parameter $\sigma\ $i.e.,$(\frac
{\omega}{4\Lambda Y}>\sigma)$, and the temperature (31) reduces to%

\begin{align}
T_{I}  &  =\underset{(\frac{\omega}{4\Lambda Y}>\sigma)\rightarrow\infty}%
{\lim}T(\omega),\nonumber\\
&  \simeq\frac{\omega}{\ln\left\{  \frac{\exp\left[  \pi\frac{\omega}%
{4\Lambda}\left(  \frac{1}{Y}+1\right)  \right]  \exp\left(  \pi\frac{\omega
}{2\Lambda}\right)  }{\exp\left[  \pi\frac{\omega}{4\Lambda}\left(  \frac
{1}{Y}-1\right)  \right]  }\right\}  },\nonumber
\end{align}

\begin{align}
&  =\frac{\omega}{\ln\left[  \exp(\frac{\pi\omega}{\Lambda})\right]  },\\
&  =\frac{\Lambda}{\pi},\nonumber
\end{align}

which smears out the $\omega-$dependence, and the spectrum results in a
isothermal radiation. In addition to this, one can immediately observe that
the above result is nothing but the $T_{Huch}$ (11), and the temperature of
the radiation becomes independent from the horizons of the charged dilaton BHs
at the high frequencies. In other words, in the high frequency regime the
charge completely loses its effectiveness on the temperature of the radiation.
Another interesting aspect of this case is that there is no possibility to
retrieve the $T_{Hch}$ (10) contary to the former studies
\cite{clement07,mazhari09} in which the high frequency regime was considered
as the master regime in obtaining the conventional Hawking temperature
($\frac{\kappa}{2\pi}$) of a BH.

\textbf{Case II: }In the low frequency regime with the conditions of
$\Lambda-2m^{2}<0$ (this case can not be obtained by shifting $m\rightarrow
im$ in the ansatz (14) of the wave function) and $\sigma>\frac{\omega
}{4\Lambda Y}$, the ratios of the hyperbolic cosines behave as%

\begin{equation}
\underset{\sigma>\frac{\omega}{4\Lambda Y}}{\lim}\frac{\cosh\left[  \pi\left(
\sigma+\frac{\omega}{4\Lambda Y}\right)  \right]  }{\cosh\left[  \pi\left(
\sigma-\frac{\omega}{4\Lambda Y}\right)  \right]  }\simeq\frac{\exp\left[
\pi\frac{\omega}{4\Lambda}\left(  1+\frac{1}{Y}\right)  \right]  }{\exp\left[
\pi\frac{\omega}{4\Lambda}\left(  1-\frac{1}{Y}\right)  \right]  },
\end{equation}
and%

\begin{equation}
\underset{\sigma>\frac{\omega}{4\Lambda Y}}{\lim}\frac{\cosh\left[  \pi\left(
\sigma+\frac{\omega}{4\Lambda}\right)  \right]  }{\cosh\left[  \pi\left(
\sigma-\frac{\omega}{4\Lambda}\right)  \right]  }\simeq\exp\left(  \pi
\frac{\omega}{2\Lambda}\right)  ,
\end{equation}
Thus for this case, the $T(\omega)$ given by (31) becomes%

\begin{align}
T_{II}  &  =\underset{\sigma>\frac{\omega}{4\Lambda Y}}{\lim}T(\omega
),\nonumber\\
&  \simeq\frac{\omega}{\ln\left\{  \frac{\exp\left[  \pi\frac{\omega}%
{4\Lambda}\left(  1+\frac{1}{Y}\right)  \right]  \exp\left(  \pi\frac{\omega
}{2\Lambda}\right)  }{\exp\left[  \pi\frac{\omega}{4\Lambda}\left(  1-\frac
{1}{Y}\right)  \right]  }\right\}  },\nonumber\\
&  =\frac{2\Lambda}{\pi}\left(  \frac{Y}{1+Y}\right)  ,\\
&  =\frac{\Lambda}{\pi}(1-\frac{r_{_{-}}}{r_{+}}),\nonumber\\
&  =\frac{M}{4\pi r_{+}}Y.\nonumber
\end{align}
which exactly matches with the result of the conventional Hawking temperature
of the charged dilaton BHs i.e., $T_{Hch}$ (10). The case II has also the
limit of $T_{Huch}$ (11) when $Y=1$ ($Q=0$).

Furthermore, we would like to represent the most interesting figures about the
spectrum temperature $T(\omega)$ (31). For this purpose, we plot $T(\omega)$
versus wave frequency $\omega$ of the 3D charged dilaton BHs with $r_{_{-}%
},r_{+}\neq0$ for both cases. Fig. 1 and Fig. 2 illustrate the thermal
behaviors of the charged dilaton BHs in the cases (I) and (II), respectively.
As it can be seen from both figures, at the high frequencies the temperature
$T(\omega)$ exhibits similar behaviors such that it approaches to $T_{Huch}$
(11) while $\omega\rightarrow\infty.$ However, in the low frequencies the
picture changes dramatically. In this regard, if we first consider the Fig. 1,
it is obvious that as $\omega\rightarrow0,$ the $T(\omega)$ tends to diverge.
Besides, Fig. 1 presents that there is no chance to have $T_{Hch}$ (10) from
the $T(\omega)$ if $\Lambda-2m^{2}\geq0.$ The minimum value of the $T(\omega
)$\ that it can drop, which is also equal to $T_{Huch}$, is higher than the
value of the $T_{Hch}$. However, when we look at the Fig. 2, for a specific
range of the low frequencies in which the conditions for the case (II) hold it
is seen that $T(\omega)$ coincides with $T_{Hch}.$ To our knowledge, the
latter result that reveals a relationship between the Hawking temperature, the
cosmological constant and the low frequencies (or long wavelengths) of the
scalar waves is completely new for the literature.

\section{Summary and conclusions}

In this paper, we have explored the frequency dependent temperature
$T(\omega)$ of the radiation spectrum for the 3D charged dilaton BHs in the
EMD$\Lambda$ theory. For this purpose, we have made use of the previously
obtained method, \textit{SCRSM}. Its related studies
\cite{clement07,mazhari09} showed us that whenever the wave equation admits an
exact solution, this method is powerful enough to obtain the Hawking
temperature of a considered BH. By using this fact, we have first obtained the
massless scalar wave equation in that geometry. After finding the exact
solution of the radial part in terms of the hypergeometric functions and using
their one of the linear transformation expressions, we have obtained the
$T(\omega).$ From here on, we have analyzed in detail the behavior of
$T(\omega)$ with various parameters. It is seen that two cases come to the
forefront during the analysis. The associated cases, which are called as "case
I" and "case II", mainly depend on the which frequency regime (high/low) is
considered, the sign of the term $\Lambda-2m^{2}$ and the domination of the
terms $\sigma$\ and $\frac{\omega}{4\Lambda Y}$ against to each other. It is
deduced from the calculations and their associated figures that in the case I
($\Lambda-2m^{2}\geq0$ and $\frac{\omega}{4\Lambda Y}>\sigma$ within the high
frequency regime) $T(\omega)$ does not yield the standard Hawking temperature
of the 3D charged dilaton BHs ($T_{Hch}$), it has only the $T_{Huch}$ limit.
On the other hand, the case II in which the frequency regime is low and
$\Lambda-2m^{2}<0$ with $\frac{\omega}{4\Lambda Y}<\sigma$ makes the
\textit{SCRSM} agree with the $T_{Hch}$. This result, by contrast with the
previous studies \cite{clement07,mazhari09}, represents that the presence of
the $\Lambda$ allows us to obtain the desired Hawking temperature $T_{Hch}$ in
the low frequency (or long wavelength) regime. The latter remark might also
play a crucial role in the detection of the charged dilaton black holes with
the cosmological constant in the future.

\end{document}